\documentclass[10pt]{article}
\usepackage{graphicx,times}
\usepackage{epsf}
\usepackage{amsmath,amssymb,latexsym,float,algorithm,algorithmic}
\usepackage{enumerate}
\usepackage{listings}
\numberwithin{equation}{section}

\newcommand{\R}{\text{\fontshape{n}\selectfont I\kern-.42exR}}

\newcommand{\1}{\text{\fontshape{n}\selectfont 1\kern-.56exl}}

\begin{document}
\title{
{\bf Speeding up Domain Wall Fermion Algorithms using QCDLAB}
}

\author{Artan Bori\c{c}i
\footnote{Invited talk given at the 'Domain Wall Fermions at Ten Years', Brookhaven National Laboratory, 15-17 March 2007}\\
        {\normalsize\it Physics Department, University of Tirana}\\
        {\normalsize\it Blvd. King Zog I, Tirana-Albania}\\
        {\normalsize\it borici@fshn.edu.al}\\
}

\date{}
\maketitle

\begin{abstract}
Simulating lattice QCD with chiral fermions and indeed using Domain Wall Fermions continues to be challenging project however large are concurrent computers. One obvious bottleneck is the slow pace of prototyping using the low level coding which prevails in most, if not all, lattice projects. Recently, we came up with a new proposal, namely QCDLAB, a high level language interface, which we believe will boost our endeavours to rapidly code lattice prototype applications in lattice QCD using MATLAB/OCTAVE language and environment. The first version of the software, QCDLAB 1.0 offers the general framework on how to achieve this goal by simulating set of the lattice Schwinger model {\tt http://phys.fshn.edu.al/qcdlab.html}. In this talk we introduce QCDLAB 1.1, which extends QCDLAB 1.0 capabilities for real world lattice computations with Wilson and Domain Wall fermions.
\end{abstract}

\vspace{14cm}
\pagebreak

\tableofcontents

\section{The challenge of Domain Wall Fermions computations}

Lattice chiral symmetry is an important ingredient for light fermion physics. In the early 1990s there were generally two seemingly different formulations of chiral fermions on the lattice: Domain Wall Fermions \cite{Ka92,FuSha95} and Overlap Fermions \cite{NaNe93,Ne98}. Introduction of Truncated Overlap Fermions \cite{Borici_TOV}, a generalisation of Domain Wall Fermions, made possible to establish the equivalence between these formulations \cite{Borici_00,Borici_qcdna3_intro}. More recently, Moebius Fermions \cite{Brower_et_al_04}, a parametric generalisation of Domain Wall Fermions, allow for more flexibility in reducing the chiral symmetry breaking effects.

Either way, the lattice chiral fermion requires the introduction of an extra dimension coupled to the four other dimensions of the standard lattice theory. Hence, Domain Wall Fermions, as a 5D problem, poses a computational challenge. Well known algorithms knowing to be working for standard fermions are no more useful and one is forced to use CGNE, which converges slowly. Therefore, it is important to search for faster algorithms. In this talk, we present QCDLAB 1.1, an algorithmic research tool for 4 and 5 dimensional fermions. The tool, a MATLAB/OCTAVE based environment, allows fast prototyping of linear algebraic computations and thus accelerates the process of finding the most efficient fermion algorithm.

\section{The philosophy behind QCDLAB}

Lattice QCD, an industrial-range computing project, is in its fourth
decade. It has basically two major computing problems: simulation of
QCD path integral and calculation of quark propagators.
Generally, these problems lead to very intensive computations and require
high-end computing platforms.

However, we wish to make a clear distinction between lattice
QCD prototyping and production codes. This is
very important in order to develop a compact and easily manageable
computing project. While this is obvious in theory,
it is less so in lattice QCD practising:
those who write lattice codes are focused primary on writing
production codes.

The code of a small project is  usually small,
runs fast, it is easy to access,
edit and debug. Can we achieve these features for a lattice prototyping code?
Or, can we modify the goals of the lattice project in order to get such
features? In our opinion, this is possible for a {\it prototyping code,
a minimal possible code
which is able to test gross features of the theory and algorithms at
shortest possible time and largest acceptable errors on a standard
computing platform}. This statement needs more explanation:

\begin{itemize}
\item[a.] Although it is hard to give sharp constraints
on the number of lines of the prototyping code, we would call 
``minimal'' that code which is no more than a few
printed pages.
\item[b.] The run time depends on computing platforms and algorithms,
and the choice of lattice action and parameters.
It looks like a great number of degrees of
freedom here, but in fact there are hardly good choices in order to reduce
the run time of a prototyping code
without giving up certain features of the theory.
Again, it is difficult to give run times.
However, a ``short'' run time should not
exceed a few minutes of wall-clock time.
\item[c.] We consider a computing platform as being ``standard''
if its cost is not too high for an academic computing project.
\item[d.] We call simulation errors to be the ``largest acceptable''
if we can distinguish clearly signal form noise
and when gross features of
the theory are not compromised by various approximations or choices.
\item[e.] Approximations should not alter basic features of the theory.
The quenched approximation, for example, should not be considered as an
acceptable approximation when studying QCD with light quarks.
\end{itemize}

A prototyping code with these characteristics should
{\it signal the rapid advance in the field},
in which case, precision lattice computations
are likely to happen in many places around the world.
Writing a minimal prototyping code is a {\it challenge of three smarts}:
smart computers, smart languages and smart algorithms.

\bigskip

QCDLAB project is based on the philosophy described above.
In the following we first describe briefly QCDLAB 1.0 and then the 1.1 version.

\section{QCDLAB 1.0}

QCDLAB is a high level language tool a collection of
MATLAB functions for the simulation of lattice Schwinger model.
It can be used as a small laboratory to test and validate algorithms.
In particular, QCDLAB 1.0
serves as an illustration of the minimal prototyping code concept.

QCDLAB 1.0 can also be used for newcomers in the field. They can learn
and practice lattice projects which are based on short codes and run
times. This offers a ``learning by doing'' method, perhaps a quickest
route into answers of many unknown practical questions concerning
lattice QCD simulations.

The next two sections describe basic algorithms
for simulation of lattice QCD and foundations of Krylov subspace methods.
Then, we present the \mbox{QCDLAB 1.0} functions followed by examples
of simple computing projects. The last section outlines the future plans
of the QCDLAB project.for lattice QCD computations. 
It is based on the MATLAB and OCTAVE language and environment.
While MATLAB is a product of The MathWorks, OCTAVE is its clone,
a free software under the terms of the GNU General Public License.

MATLAB/OCTAVE is a technical computing environment integrating numerical
computation and graphics in one place, where problems and solutions look
very similar and sometimes almost the same as they are written mathematically.
Main features of MATLAB/OCTAVE are:

\begin{itemize}
\item Vast Build-in mathematical and linear algebra functions.
\item Many functions form Blas, Lapack, Minpack, etc. libraries.
\item State-of-the-art algorithms.
\item Interpreted language.
\item Dynamically loaded modules from other languages like C/C++, FORTRAN.
\end{itemize}

QCDLAB 1.0 is 
\begin{itemize}
\item General functions:
\begin{itemize}
\item Solvers: {\tt BiCGg5, BiCGstab, CG, CGNE, FOM, GMRES, Lanczos, SCG, SUMR}
\item Data processing: {\tt Autocorel, Binning}.
\end{itemize}
\item Specialised functions for the Schwinger model:
\begin{itemize}
\item Simulation: {\tt HMC\_W, HMC\_KS, Force\_W, Force\_KS}
\item Operators: {\tt Dirac\_KS, Dirac\_r, Dirac\_W, cdot5}
\item Measurements: {\tt wloop}
\end{itemize}
\end{itemize}a collection of
functions for the simulation of lattice Schwinger model.
It can be used as a small laboratory to test and validate algorithms.
In particular, QCDLAB 1.0
serves as an illustration of the minimal prototyping code concept.

QCDLAB 1.0 can also be used for newcomers in the field. They can learn
and practice lattice projects which are based on short codes and run
times. This offers a ``learning by doing'' method, perhaps a quickest
route into answers of many unknown practical questions concerning
lattice QCD simulations. It contains the following MATLAB/OCTAVE functions:

\vspace{0.5cm}
{\tt
\begin{tabular}{lllll}
Autocorel  & BiCGg5    & BiCGstab    & Binning   & cdot5\\
CG         & CGNE      & Dirac\_KS   & Dirac\_r  & Dirac\_W\\
FOM        & Force\_KS & Force\_W    & GMRES     & HMC\_KS\\
HMC\_W     & Lanczos   & SCG         & SUMR      & wloop
\end{tabular}
}
\vspace{0.5cm}

The functions can be grouped as in the following:

\begin{itemize}
\item General functions:
\begin{itemize}
\item Solvers: {\tt BiCGg5, BiCGstab, CG, CGNE, FOM, GMRES, Lanczos, SCG, SUMR}
\item Data processing: {\tt Autocorel, Binning}.
\end{itemize}
\item Specialised functions for the Schwinger model:
\begin{itemize}
\item Simulation: {\tt HMC\_W, HMC\_KS, Force\_W, Force\_KS}
\item Operators: {\tt Dirac\_KS, Dirac\_r, Dirac\_W, cdot5}
\item Measurements: {\tt wloop}
\end{itemize}
\end{itemize}

In order to get started with QCDLAB 1.0 one can run the following three projects:
\begin{itemize}
\item For Matlab: {\tt MProject1, MProject2, MProject3}
\item For Octave: {\tt OProject1, OProject2, OProject3}
\end{itemize}

The user is required to download these m-files to the working directory of MATLAB/OCTAVE
and then type the corresponding project names in the MATLAB/OCTAVE environment.  
The first project is a simulation project, the second one is a linear system
and eigenvalue solver project, whereas the third one is a
linear system and eigenvalue solver project for chiral fermions.

For more information on QCDLAB 1.0 the reader is referred to the complete documentation available
at the project web page\\
{\tt http://phys.fshn.edu.al/qcdlab.html}.

\section{QCDLAB 1.1}

As stated in the first section, the goal of the QCDLAB project is to create an
algorithmic prototyping environment for lattice QCD computations. This goal has no ending station, but rather it is a process which will enhance the computing capabilities as the time goes on. The first step in this direction is the 1.1 version. It offers new functionality for 4D and 5D computations at the linear system and eigenvalue solver level. The new functions are:

\begin{verbatim}
cdot5             Dirac4             Initialise_Dirac_W     InversePower
Mult_Dirac_W      Mult_Dirac_W_H     Mult_DWF               Mult_DWF_H
mult_gamma5       P5minus            P5plus                 PowerMethod
\end{verbatim}

There are two ways to implement the Wilson operator:
\begin{itemize}
\item Classical way of matrix-vector multiplication using {\tt Mult\_Dirac\_W, Mult\_Dirac\_W\_H}
\item Creation of a sparse matrix using {\tt Dirac4}
\end{itemize}

In the first case one needs to initialise the Wilson operator using the {\tt Initialise\_Dirac\_W} function and to hack inversion solvers of QCDLAB 1.0 in the place where the multiplication with {\tt A} takes place, eg.

{\tt A*x} $\rightarrow$ {\tt Mult\_Dirac\_W(x)}

The same comment is for the Domain Wall Fermion matrix-vector functions {\tt Mult\_DWF, Mult\_DWF\_W}. Note that the 1.0 version of the {\tt cdot5} function is now redefined for the 4D fermion theory and is included in the 1.1 version. Other useful functions are multiplication of a 4D-vector by $\gamma_5$: {\tt mult\_gamma5}, the chiral projection functions applied to 4D-vectors: {\tt P5minus, P5plus}.

\subsection*{Gauge field data structure}

Both {\tt Dirac4} and {\tt Initialise\_Dirac\_W} need the gauge field, which must be supplied as a set of four matrices: {\tt u1}, {\tt u2}, {\tt u3}, {\tt u4}. Each gauge field component is a $9N\times 2$ matrix, where $N$ is the total number of lattice sites. The first column is the real part and the second the imaginary part of the particular SU(3) matrix element. Note that, if reshaped, the most inner dimensions of gauge field component are $3\times 3$ matrices, i.e.
\begin{verbatim}
reshape{u1,3,3,N,2}
\end{verbatim}
The user should care to organise the gauge field data in this way for the QCDLAB functions to work as required.

\subsection*{Sparse Wilson matrices}

In case one want to create a sparse Wilson matrix one uses the {\tt Dirac4} function:
\begin{verbatim}
function A=Dirac4(u1,u2,u3,u4);
% Constructs Wilson-Dirac operator
mass=0; N1=8; N2=8; N3=8; N4=16; N=N1*N2*N3*N4;
% Gamma matrices
gamma1 = [0, 0, 0,-i;  0, 0,-i, 0;  0, i, 0, 0;  i, 0, 0, 0];
gamma2 = [0, 0, 0,-1;  0, 0, 1, 0;  0, 1, 0, 0; -1, 0, 0, 0];
gamma3 = [0, 0,-i, 0;  0, 0, 0, i;  i, 0, 0, 0;  0,-i, 0, 0];
gamma4 = [0, 0,-1, 0;  0, 0, 0,-1; -1, 0, 0, 0;  0,-1, 0, 0];
% Projection operators
P1_plus = eye(4)+gamma1; P1_minus=eye(4)-gamma1;
P2_plus = eye(4)+gamma2; P2_minus=eye(4)-gamma2;
P3_plus = eye(4)+gamma3; P3_minus=eye(4)-gamma3;
P4_plus = eye(4)+gamma4; P4_minus=eye(4)-gamma4;
% Shift operators
p1=[N1,1:N1-1]; p2=[N2,1:N2-1]; p3=[N3,1:N3-1]; p4=[N4,1:N4-1];
I1=speye(N1); I2=speye(N2); I3=speye(N3); I4=speye(N4);
T1=I1(:,p1); T2=I2(:,p2); T3=I3(:,p3); T4=I4(:,p4);
E1=spkron(I4,spkron(I3,spkron(I2,spkron(T1,speye(3)))));
E2=spkron(I4,spkron(I3,spkron(T2,spkron(I1,speye(3)))));
E3=spkron(I4,spkron(T3,spkron(I2,spkron(I1,speye(3)))));
E4=spkron(T4,spkron(I3,spkron(I2,spkron(I1,speye(3)))));
% Gauge Field configuration {u1, u2, u3, u4}: 9*N by 2 matrices
I_N=speye(N);
[I,J]=spfind(spkron(I_N,ones(3)));
U1=sparse(I,J,u1(:,1)+i*u1(:,2),3*N,3*N);
U2=sparse(I,J,u2(:,1)+i*u2(:,2),3*N,3*N);
U3=sparse(I,J,u3(:,1)+i*u3(:,2),3*N,3*N);
U4=sparse(I,J,u4(:,1)+i*u4(:,2),3*N,3*N);
% Upper triangular
U=spkron(P1_minus,U1*E1)+spkron(P2_minus,U2*E2)+spkron(P3_minus,U3*E3)+spkron(P4_minus,U4*E4);
% Lower triangular
L=spkron(P1_plus ,U1*E1)+spkron(P2_plus ,U2*E2)+spkron(P3_plus ,U3*E3)+spkron(P4_plus ,U4*E4);
%M=U+L';
A=(mass+4)*speye(12*N)-0.5*(U+L');
% Copyright (C) 2006-2007 Artan Borici.
% This program is a free software licensed under the terms of the GNU General Public License
\end{verbatim}

\subsection*{Eigenvalue solvers}

The 1.1 version comes with two eigenvalue solvers: {\tt PowerMethod, InversePower}, which are implementations of the methods with the same name. They can be used for the Hermitian eigenvalue problems. For example, if one would like to compute the smallest eigenvalue of the Hermitian Wilson operator one can use the {\tt InversePower} function:
\begin{verbatim}
function [v,lambda,rr]=InversePower(b,x0,tol,nmax);
% Inverse power method for the Hermitian Wilson operator
v=b/norm(b); rr=[];
while 1,
  u=bicg5(v,x0,1e-6,1000);
  u=mult_gamma5(u);
  lambda=v'*u;
  r=v-u/lambda;
  rnorm=norm(r); rr=[rr;rnorm];
  if rnorm<tol, break, end
  v=u/norm(u);
end
% Copyright (C) 2006-2007 Artan Borici.
% This program is a free software licensed under the terms of the GNU General Public License
\end{verbatim}

\subsection*{Domain Wall Fermion operator}

The {\tt Mult\_DWF} implements the Domain Wall Fermion operator
$$
{\mathcal M} =
\begin{pmatrix}
~\1-D_W          & ~~~~~P_+ &                & -mP_- \\
~~~~~~P_-     & ~\1-D_W      & \ddots         &
  \\
                & \ddots      & \ddots         & P_+
  \\
-mP_+ &             & P_-    & ~\1-D_W
  \\
\end{pmatrix}
$$
applied to a vector:
\begin{verbatim}
function y=Mult_DWF(x,N5);
% Multiplies a vector by the Domain Wall Fermion matrix
global N mass_dwf

x=reshape(x,12*N,N5);
%
y(:,1)=x(:,1)-Mult_Dirac_W(x(:,1))+P5plus(x(:,2))-mass_dwf*P5minus(x(:,N5));
for j5=2:N5-1;
  y(:,j5)=x(:,j5)-Mult_Dirac_W(x(:,j5))+P5plus(x(:,j5+1))+P5minus(x(:,j5-1));
end
y(:,N5)=x(:,N5)-Mult_Dirac_W(x(:,N5))-mass_dwf*P5plus(x(:,1))+P5minus(x(:,N5-1));

x=reshape(x,12*N*N5,1);
y=reshape(y,12*N*N5,1);
%
% Copyright (C) 2006-2007 Artan Borici.
% This program is a free software licensed under the terms of the GNU General Public License
\end{verbatim}

\subsection*{Acknowledgements}

The author wishes to thank Tom Blum and Amarjit Soni for the invitation and the kind hospitality at BNL as well as Stefan Sint for useful discussion on possible extensions of QCDLAB's Dirac operators with non-trival boundary conditions.


\begin{thebibliography}{10}
 
\bibitem[Kaplan 1992]{Ka92}
        D.B. Kaplan,
        {\it A Method for Simulating Chiral Fermions on the Lattice}
        Phys. Lett. B 228 (1992) 342.

\bibitem[Furman and Shamir 1995]{FuSha95}
        V. Furman, Y. Shamir,
        {\it Axial symmetries in lattice QCD with Kaplan fermions},
        Nucl. Phys. B439 (1995) 54-78

\bibitem[Narayanan and Neuberger 1993]{NaNe93}
        R. Narayanan, H. Neuberger,
        {\it Infinitely many regulator fields for chiral fermions},
        Phys. Lett. B 302 (1993) 62,
        {\it A construction of lattice chiral gauge theories},
        Nucl. Phys. B 443 (1995) 305.

\bibitem[Neuberger 1998]{Ne98}
        H. Neuberger,
        {\it Exactly massless quarks on the lattice},
        Phys. Lett. B 417 (1998) 141

\bibitem[Bori\c{c}i 2000c]{Borici_TOV}
         A. Bori\c{c}i,
         {\it Truncated Overlap Fermions},
         Nucl. Phys. Proc. Suppl. 83 (2000) 771-773

\bibitem[Bori\c{c}i 2000]{Borici_00}
        A. Bori\c{c}i,
        {\it Truncated Overlap Fermions: the link between
        Overlap and Domain Wall Fermions},
        in V. Mitrjushkin and G. Schierholz (edts.),
        Lattice Fermions and Structure of the Vacuum,
        Kluwer Academic Publishers, 2000.

\bibitem[Bori\c{c}i 2005a]{Borici_qcdna3_intro}
        A. Bori\c{c}i,
        {\it Computational methods for the fermion determinant
        and the link between overlap and domain wall fermions},
        in QCD and Numerical Analysis III, ed. Bori\c{c}i
        {\it et al}, Springer 2005.

\bibitem[Brower {\it et al} 2004]{Brower_et_al_04}
         R.C. Brower, H. Neff, K. Orginos,
         {\it Mobius Fermions: Improved Domain Wall Chiral Fermions},
         \texttt{hep-lat/0409118}

\end{thebibliography}
\end{document}